\documentclass[preprint,showpacs,preprintnumbers,amsmath,amssymb]{revtex4}

\usepackage{graphicx}
\usepackage{dcolumn}
\usepackage{bm}

\begin{document}

\preprint{APS/123-QED}

\title{Spin-orbit induced non-collinear spin structure \\ 
in deposited transition metal clusters}

\author{S.~Mankovsky, S.~Bornemann, J.~Min\'ar, S.~Polesya, H.~Ebert}
\affiliation{%
Department of Chem.\ and Biochem./Phys.\ Chem., LMU Munich,\\
Butenandtstrasse 11, D-81377 Munich, Germany 
}%

\author{J.~B.~ Staunton}
\affiliation{%
Department of Physics, University of Warwick, Coventry CV4 7AL,
           United Kingdom
}%

\author{A. I. Lichtenstein}
\affiliation{%
 Institut f\"ur Theoretische Physik,
Universit\"at Hamburg, D-20355 Hamburg, Germany
}%

\date{\today}

\begin{abstract}
The influence of the spin-orbit coupling on the magnetic structure of
deposited transition metal nanostructure systems has been studied by
fully relativistic electronic structure calculations.
 The interplay of exchange coupling and magnetic anisotropy
 was monitored by studying the corresponding magnetic torque calculated
 within ab-initio and model approaches. 
It is found that a spin-orbit induced Dzyaloshinski-Moriya interaction
can stabilise a non-collinear spin structure even if there is a
pronounced isotropic ferromagnetic exchange interaction between the 
magnetic atoms. 
\end{abstract}

\pacs{Valid PACS appear here}
\maketitle



As the complexity and technological applicability of nanostructured 
magnetic materials grows it has become important to develop a reliable
quantitative theoretical framework in which to understand them. In
principle this is available from relativistic density functional theory
(RDFT) \cite{MV79}. The magnetic behaviour of complex magnetic systems
is described effectively in terms of 'local moments', even in metallic
systems, which fluctuate as the temperature is increased from 0K. Within
the effective single electron picture of DFT, the moments arise from
local magnetic fields associated with the different atomic sites which
affect the electronic motions and are self-consistently maintained by
them. The spin-orbit coupling (SOC) effects on the electronic structure
determine the magnetic anisotropy linking the magnetic 
and spatial structure of a material. 
%
%
As demonstrated by Staunton et al. \cite{SSB+06} and others, magnetic
torque calculations enable magnetic anisotropy to be studied reliably
and magnetic structures to be determined. We have recently developed a
method to study ab-initio the magnetic structures of complex
nanostructures using this approach \cite{BMS+07}. It is based on the
framework of relativistic density functional theory using the local
spin density approximation (LSDA) for exchange and correlation
effects~\cite{VWN80}. The electronic structure was determined in a 
fully relativistic way from the basis of the Dirac equation for spin-polarized
potentials which is solved using the Korringa-Kohn-Rostoker (KKR) multiple
scattering formalism \cite{Ebe00}.

In this letter we set out to understand the relativistic effects on the
magnetic structure of nanoclusters in simpler terms of a model `moment'
Hamiltonian. In this context the mapping of energetic properties
obtained from first principles calculations of complex magnetic systems
onto a Heisenberg Hamiltonian has proved over recent years to be a very
robust and successful scheme \cite{LKAG87,USPW03}. Moreover for many
systems the results concur with those from an ab-initio `disordered
local moment' theory in which no prior mapping to a Heisenberg system is
assumed \cite{GPS+85,SORG04}.  An extensively used approach to calculate
the isotropic exchange interaction parameter $J_{ij}$ for two magnetic
moments on sites $i$ and $j$ for use in the classical 
Heisenberg Hamiltonian was worked out by Lichtenstein et
al. \cite{LKAG87} using perturbation theory and the so-called Lloyd
formula. A corresponding fully relativistic approach was later introduced by
Udvardi et al. \cite{USPW03} that produces an exchange
interaction tensor $\underline{\underline{J}}_{ij}$ for use in a prescribed 
extended Heisenberg Hamiltonian. 
This scheme generates in particular a Dzyaloshinski-Moriya (DM)-type
interaction \cite{Dzy58,Mor60}, 
which may explain many interesting phenomena including the magnetic ground
state configuration of nanostructures \cite{ALU+08,VUWW07} as well as
magnetic thin films  \cite{BHB+07,FBV+08}.
Here we explore what form the effective moment Hamiltonian should have
by a detailed study of its building blocks.
With an investigation of Fe$_2$,
Co$_2$ and Ni$_2$ dimers deposited on Pt(111) as most simple cluster
examples we identify strong
DM-type interactions as well as an additional 
substrate-generated effect in Ni$_2$ (for calculational details see \cite{BMS+07}). The conclusions of these results
can then be applied straightforwardly to larger clusters and nanostructures.

We start by considering the magnetic torque vector $\vec{T}_i^{(\hat{e}_i)}$
acting on an atomic magnetic moment on a site $i$ ($i=1$ or $2$) and
aligned along 
direction  $\hat{e}_i$. The torque vector is defined in terms of the change in 
energy $E(\{\hat{e}_k\})=E(\hat{e}_1, \hat{e}_2) $ of the system when
changing the orientation of the magnetic moment, $\hat{e}_i$ on site
$i$, $\vec{T}_i^{(\hat{e}_i)}= - \partial 
E(\{\hat{e}_k\})/\partial \hat{e}_i$. 
The component $T^{(\hat{e}_i)}_{i,\hat{u}} 
= - (\partial E(\{\hat{e}_k\})/\partial \hat{e}_i)\cdot (\hat{u} \times
\hat{e}_i)$ of $\vec{T}_i^{(\hat{e}_i)}$ with respect to the axis $\hat{u}$
can be determined from first-principles 
using an expression derived by  Staunton et al.
\cite{SSB+06}. Following on from this the derivative ${\partial^2
  E}/{\partial\hat{e}_i \partial\hat{e}_j}$ describing  the change in
energy upon changing the orientation of two magnetic moments on sites $i$ and
$j$ can also be obtained \cite{LKAG87,USPW03}. By making use of the rigid spin
approximation (RSA) \cite{AKH+96} this approach leads to a fitting of
the magnetic energy landscape $E(\{\hat{e}_k\})$ at low temperatures
obtained from first principles calculations onto a Heisenberg model
Hamiltonian. With SOC included an anisotropy in the exchange interaction
may occur. We then use these quantities in an extended classical
Heisenberg Hamiltonian for the `spins',$\{\hat{e}_k\}$ of the following form,
(e.g.\cite{AKL97, USPW03}):
\begin{eqnarray}
 H &=& 
-\frac{1}{2}\sum_{i,j (i \neq j)} J_{ij}
 \hat{e}_i \cdot \hat{e}_j 
-\frac{1}{2} \sum_{i,j (i \neq j)}
 \hat{e}_i\underline{\underline{J}}^S_{ij} \hat{e}_j 
\nonumber \\
& & - \frac{1}{2}
 \sum_{i,j (i \neq j)} \vec{D}_{ij}\cdot [\hat{e}_i\times
 \hat{e}_j] 
 +\sum_{i}  K_i (\hat{e}_i)\;.  
\label{Hspin_2}
\end{eqnarray}
Here the exchange interaction  tensor $\underline{\underline{J}}_{ij}$
has been split into its conventional isotropic part $J_{ij}$, 
its traceless symmetric
part $\underline{\underline{J}}^S_{ij}$ and its anti-symmetric part
$\underline{\underline{J}}^A_{ij}$. We assume that the latter one is represented in
terms of the 
DM vector $\vec{D}_{ij}$ with
$D^{\gamma}_{ij} = \epsilon^{\alpha\beta\gamma} \frac{J^{\alpha\beta}_{ij} -
  J^{\beta\alpha}_{ij}}{2}$ ($\epsilon^{\alpha\beta\gamma}$ is the Levi-Civita
symbol). Finally, the anisotropy constants  $K_i(\hat{e}_i)$ account for
 the so-called on-site anisotropy energy
associated with each  individual moment oriented along $\hat{e}_i$. From our ab-initio
calculation of the torques $T^{(\hat{e}_i)}_{i,\hat{u}}$ we can test whether such a model
Hamiltonian is justified and also find the values of the  $J_{ij}$, $K_i$, $\underline{\underline{J}}^S_{ij}$, 
and $\vec{D}_{ij}$ parameters.

For the model Eq.~(\ref{Hspin_2})
the rate of change in energy when a moment on site $i$ is rotated about an axis $\hat{u}$ 
can be partitioned into contributions from different terms of Heisenberg
Hamiltonian $T^{(\hat{e}_i)}_{i,\hat{u}}=
T^{\rm iso}_{i,\hat{u}} + T^{\rm S}_{i,\hat{u}}+ T^{\rm DM}_{i,\hat{u}} +
T^{K}_{i,\hat{u}}$. The contribution to the torque from the DM coupling
is given by: 
\begin{eqnarray}
T^{\rm DM}_{i,\hat{u}}  &=& 
\sum_{j \neq i} ( \vec{D}_{ij}\cdot \hat u) (\hat{e}_i \cdot \hat{e}_j)  
 - \sum_{j\neq i} ( \vec{D}_{ij}\cdot \hat{e}_i) (\hat u \cdot \hat{e}_j) \;.
\label{Torque_DM}
\end{eqnarray}
The last term derives from the single site anisotropy term, i.e. $T^{K}_{i,\hat{u}}
= \frac{\partial K_i (\hat{e}_i)}{\partial \hat{e}_i}\cdot (\hat{u} \times \hat{e}_i)$.
By focussing on nanoclusters of collinear `spin' arrangements, the magnetically anisotropic
terms can be determined. 
Note that the last term in Eq.~(\ref{Torque_DM}) does not contribute
to the torque in the case of a collinear magnetic structure. 
Also the sum of the DM
contributions $T^{\rm DM}_{i,\hat{u}}$ from all sites in a nanocluster to the total torque vanishes in this case.
On the other hand, the anisotropy of the exchange interaction,
represented by the symmetric tensor $\underline{\underline{J}}^S_{ij}$
gives rise to a finite torque hence 
contributing to the total magnetic anisotropy energy of the nanocluster. 

\begin{figure}
\includegraphics[width=0.25\textwidth,angle=0,clip]{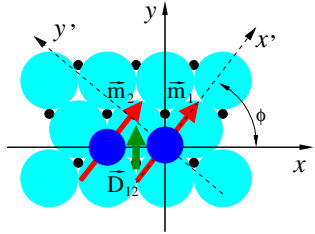}
\caption{\label{fig:geom} Magnetic configuration of the
  transition metal dimers deposited on a Pt(111) substrate. The large
  (small) spheres present Pt surface (subsurface) atoms. The medium size
  spheres represent the dimer atoms with the projection of their
  magnetic moments onto the surface ($xy$-plane) represented by arrows. The
projection of the DM-vector $\vec{D}_{12}$ onto the surface is
represented by a short arrow.}
\end{figure}

We now see how well our ab-initio
calculations of deposited Fe, Co and Ni dimers, using our first-principles
multiple scattering formalism, fit such model torques. Firstly we find that the dimers, when constrained to be magnetically collinear exhibit a pronounced out-of-plane magnetic anisotropy \cite{GRV+03}.
The collinear magnetic moments $\hat{e}_1$ and $\hat{e}_2$  have been assumed to be
orientated at an angle to the surface
normal ($z$-axis) and are expressed in terms of polar angles $\theta$ and $\phi$, as $\hat{e}_{1}=\hat{e}_2=
(\sin \theta \cos \phi,\sin \theta \sin \phi, \cos \theta)$. 
Fig.~\ref{fig:geom} shows the atomic configuration together the
projection of the 
moments onto the surface ($xy$-plane). In parallel to the fixed frame of
reference ($x, y, z$) we use a second one 
($x', y', z'$) rotated by
$\phi$ with respect to the fixed one with $\hat{z} = \hat{z}'$ (see Fig. \ref{fig:geom}).
In all calculations the torque $T^{\hat{e}_i}_{i,\hat{u}}$ is taken
around the $y'$ axis, i.e. $\hat{u} = \hat{y}'=
(- \sin \phi, \cos \phi, 0)$, for $\theta$ fixed to $\pi/4$ as a
function of the azimuthal angle $\phi$.
From the model Eq.~(\ref{Hspin_2}) an expansion of the anisotropy energy term $K(\hat{e}_i)$ in terms of $l=2$
spherical harmonics, gives the single site contribution to
the torque in present configuration to be 
$T^{K} =  -2[K_{2,1} + K_{2,2}\cos(2\phi)+ K'_{2,2}\sin(2\phi)]$ whereas $T^{\rm DM}_{i,\hat{u}}$ deduced from the
model has a $\cos(\phi)$ ($\sin(\phi)$) variation when $\vec{D}_{1,2}$ lies in the $yz$ ($xz$) plane.
%
%

\begin{figure}
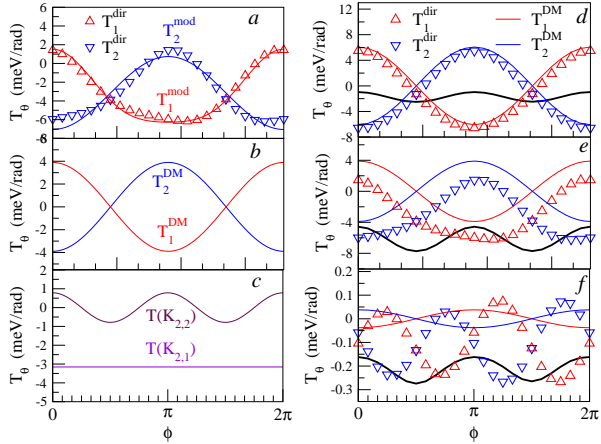

\includegraphics[width=0.23\textwidth,angle=0,clip]{Co2_DM_2.eps}
\includegraphics[width=0.24\textwidth,angle=0,clip]{FeCoNi_2_T_DM_2.eps}
\caption{\label{fig:Co_MCA} 
 Magnetic torque $T_{i}$ for dimers  on
  Pt(111) for $\theta = \pi/4$ as a function of the azimuthal angle
  $\phi$. Left: Torque for Co$_2$/Pt(111): {\it a}) results $T^{\rm dir}_{i}$
  of direct calculations (up and  down  triangles) compared 
  with results $T^{\rm mod}_{i}$ obtained on the basis of the Heisenberg-type
  model Hamiltonian (Eq.~(\ref{Hspin_2})); {\it b}) $T^{\rm DM}_i$ contribution
  to $T^{\rm mod}_{i}$ due to the DM interaction
  (Eq.~(\ref{Torque_DM})); {\it c}) contribution $T^{K}_i$ to  $T^{\rm
    mod}_{i}$ due to the on-site anisotropy. 
  Right: Magnetic torque for the two atoms of Fe$_2$ ({\it d}),  Co$_2$ ({\it e})
  and Ni$_2$ ({\it f}). The results calculated directly are presented by
  up ($T^{\rm dir}_{1}$) and  down ($T^{\rm dir}_{2}$) triangles and
  their sum by thick solid line ($T^{\rm dir}_{\rm tot}$ = $T^{\rm dir}_{1}$ + 
  $T^{\rm dir}_{2}$). The thin solid lines give
  the DM contributions according to Eq.~(\ref{Torque_DM}).
} 
\end{figure}

In Fig.~\ref{fig:Co_MCA} the ab-initio results for the torque
$T_i^{\rm dir}$ are  compared with those deduced from the model
Eq.~(\ref{Hspin_2}). 
As one notes, the torques for the two atoms are different but are related
with respect to their $\phi$-dependence according to the C$_s$ symmetry of
the system. 
On comparing  the torque of
the two Co atoms of Co$_2$/Pt(111) obtained directly from the electronic
structure calculations 
with those from the model (see
Fig.~\ref{fig:Co_MCA}{\it a-c}) we find the symmetric part of the
exchange interaction tensor $T_i^S$ to be  negligible. The contribution
$T^{\rm DM}_i$ due to the DM interaction Eq.~(\ref{Torque_DM}) is shown in
Fig.~\ref{fig:Co_MCA}{\it b}. Clearly, $T^{\rm DM}_1$ and
$T^{\rm DM}_2$ vary with $\cos \phi$ and are opposite in sign. In accordance
with the C$_s$ symmetry, the DM vector
$\vec{D}_{12}$ lies in the $yz$ plane (Fig.~\ref{fig:geom}). For the
contribution $T^{K}_i$ due to the on-site anisotropy one finds
 $K'_{2,2}$ to be very small and the
dominating terms $K_{2,1}$ and $K_{2,2}$ to be practically the same for both atomic sites  leading to $T^{K}_1 \approx T^{K}_2$.

The contribution to $T^{K}_i$ connected
with $K_{2,1}$ does not depend on $\phi$, while that connected with
$K_{2,2}$ varies with $\cos(2\phi)$ (Fig.~\ref{fig:Co_MCA}{\it c}). As one can see, the torque $T^{\rm mod}_i$ derived
from the model Hamiltonian  reproduces the
results $T^{\rm dir}_i$ calculated directly rather well. The remaining
deviation
is primarily due to the limitations of the model
Hamiltonian with respect to the dependency of magnetic energy on
the magnetic moment orientations $E(\{\hat{e}_k\})$. From the
decomposition of the $T_i$'s  via the model Hamiltonian it becomes  clear that
its $\phi$-dependence is dominated by the DM-contribution while the
$K_{2,2}$-contribution gives rise to a minor additional modulation. 
Owing to the large negative values of $T_1(K_{2,1})$ and $T_2(K_{2,1})$ an out-of-plane anisotropy
results for the total system. This can be seen also from
Fig.~\ref{fig:Co_MCA}{\it e} where the total torque 
$T_{\rm tot}^{\rm dir} = T_{1}^{\rm dir} + T_{2}^{\rm dir}$ for Co$_{2}$ on
Pt(111) is shown together with the individual contributions $T_{i}^{\rm
  dir}$ and the corresponding DM terms $T_{i}^{\rm DM}$. 
%
%
Once more one sees that the
$\phi$-dependence of the individual torques is determined by
$T_{i}^{\rm DM}$ while that of the total torque $T_{\rm tot}^{\rm dir}$ is
set by the $K_{2,2}$ on-site anisotropy terms. This also holds for
Fe$_2$ (Fig.~\ref{fig:Co_MCA}{\it d}) for which the DM terms are
even more dominant, i.e. the $\phi$-dependence of the individual torques
are nearly exclusively due to  $T_{i}^{\rm DM}$.

The situation is noticeably
different for Ni$_{2}$ on Pt(111) (Fig.~\ref{fig:Co_MCA}{\it f}) for
which the DM terms give only minor contributions to the individual
torques $T_1$ and $T_2$. In contrast to Fe$_2$ and Co$_2$ the difference
between the torques of different atoms cannot be attributed to DM
coupling. The period of oscillations of the torques as functions of
$\phi$ are different compared to those of the torques created by DM coupling
varying as $\sin(2\phi)$. Clearly an effect which is not contained in
the model Hamiltonian Eq.~(\ref{Hspin_2})) is evident and must  
derive from interactions of the Ni moments with those induced in the
Pt substrate. As with a DM-type interaction it produces torques  such
that $T_1= -T_2$ but its $\sin(2\phi)$ variation would arise if a term
of the form $(\vec{A} \cdot \hat{e}_1) (\vec{B} \cdot \hat{e}_1)
-(\vec{A} \cdot \hat{e}_2)  \cdot (\vec{B} \cdot \hat{e}_2)$ were added
to Eq.~(\ref{Hspin_2}), where $\vec{A}$ lies along $x$ and $\vec{B}$
along $y$.  These Ni$_2$ dimer results indicate that an effective
Heisenberg model must be used with caution for systems where the
magnetic structure of a nanocluster is strongly influenced by the spin 
polarisability of the substrate. 

For the Co$_2$ and Fe$_2$ dimers, however, the model works very well. 
For our chosen geometry $\theta = \pi/4$ and $\phi = 0$ and 
using the symmetry properties of the elements of the exchange tensor, one
finds for Co$_2$/Pt(111) with $T_{1}^K  \approx T_{2}^K$ the total
torque  $T = T_{1} + T_{2} = -(J^{Szz}_{12} - J^{Sxx}_{12}) + 2T^{K}_1$.
Finally, the magnetic anisotropy energy (MAE) of the dimer being the difference
in energy when the  magnetic moments are both oriented along $\hat{e}_b$
and $\hat{e}_a$, is given by the
integral $-\int_{\hat{e}_a}^{\hat{e}_b}(\vec{T}_1(\hat{e})+
\vec{T}_2(\hat{e})) d\hat{e}$~\cite{BMS+07}.  
Obviously this has no contribution from the DM interaction. 
For Co$_2$/Pt(111) we find the exchange parameters $J_{12}^{xx}$,
$J_{12}^{yy}$ and $J_{12}^{zz}$ to be nearly the same implying that the total
MAE of the dimer is nearly exclusively due to the on-site
anisotropy. The values $K_{2,1} = 1.5$ meV and $K_{2,2} = 0.39$ meV for
Co$_2$/Pt(111) lead as mentioned above to a pronounced out-of plane
anisotropy, i.e. in the ground state the total magnetisation points along the
surface normal.

Taking the difference between the individual torques one arrives at the
relation $D^y_{12} = \frac{T_{1} - T_{2}}{2} $
allowing $D^y_{12}$ to be deduced directly from the ab-initio torque
$T^{\rm dir}_i$ 
calculations. Table \ref{one} shows the
corresponding  results for dimers Fe$_2$ and Co$_2$ on Pt(111)
in comparison with data derived from a mapping to our model Hamiltonian,
Eq.~(\ref{Hspin_2}).

\begin{table}\caption{\label{one} Components $D^{\alpha}_{12}$ of the DM
    vector $\vec{D}_{ij}$, the isotropic exchange constant $J_{ij}$
    (in meV) and the tilt angle $\alpha$ (in degrees) (see text) for the
    dimers Fe$_2$ and Co$_2$ on Pt(111). The data 
    labelled {\it direct} have been obtained from the direct first principles
    calculations of the torque. The
    data below have been obtained by mapping first principles results
    onto the model Hamiltonian, Eq.~(\ref{Hspin_2}).
 }
\begin{center}
\begin{tabular}{l||c|c|c|c|c|c}
\hline
 &\hspace{0.2mm}  direct \hspace{0.2mm}&
 \multicolumn{4}{c|}{model} & \\
\hline
 \hspace{0mm}    &\hspace{0.1mm} $D_{12}^y$ \hspace{0.1mm}&\hspace{0.1mm} $D_{12}^x$\hspace{0.1mm} &\hspace{0.1mm} $D_{12}^y$\hspace{0.1mm} &\hspace{0.1mm} $D_{12}^z$\hspace{0.1mm} &
 $J_{12}$ &\hspace{0.6mm} $\alpha$
 \hspace{5mm}   \\
\hline
 Fe$_2$ &6.04 & 0.00 & 6.07 & -3.34 & 138.0 & 2.52 \\ 
\hline
Co$_2$  &3.69 & 0.00 & 3.89 & -3.84 & 108.0 & 2.07  \\ 
\hline
Ni$_2$  &-0.02 & 0.00 &-0.04 &-0.24 & 30.4 & 0.07  \\ 
\hline
\end{tabular}
\end{center}
\end{table}


The closeness of the results justifies once
more the use of the model Hamiltonian for Fe$_2$ and Co$_2$. In addition one notes that
$D^y_{12}$ has an appreciable value compared to the
isotropic exchange constant $J_{12}$.
Fixing the azimuthal angle $\phi$ to be $\pi/2$ and performing similar
steps one finds $D_{12}^x$ to be $0$. This is also in line with
the $C_s$ symmetry of the investigated dimer systems (see
Fig.~\ref{fig:geom}). $D_{12}^z$, on the other hand, may take a
non-zero value and is found to be comparable to $D_{12}^y$ (see Table \ref{one}).

Thus, the above analysis shows that the torques
$T_{1}$ and $T_2$  may differ even if the
total torque is zero, i.e. if the moments are aligned collinearly along
the easy axis (surface normal). The difference between these torques is
caused exclusively by the $D_{12}^y$ terms leading to a rotation around
the $y$-axis. Minimizing the magnetic 
energy $E(\{\hat{e}_i\})$ of the 2 atom clusters leads to an outward
tilting of the moments by 
an angle $\alpha$ given by $\alpha =$ atan $(D^y_{12}/J_{12})$ with
$J_{12}$ being the isotropic exchange coupling constant.
The corresponding results given in Table \ref{one} show that
the DM interaction causes the deposited Fe$_2$ and Co$_2$ dimers to have an appreciable
deviation from collinear configurations in spite of the pronounced
ferromagnetic exchange coupling given by $J_{12}$. This effect of
SOC is completely in line with the findings of Sandratskii and K\"ubler
 \cite{San98} for bulk systems. 

The substrate clearly plays a crucial role in the DM
interaction. Firstly, hybridisation with the substrate
breaks the inversion symmetry for the dimer leading to a non-zero DM
vector. This symmetry effect is also confirmed
by test calculations on free dimers and on dimers embedded in a bulk Pt
matrix. Secondly, the hybridisation with the substrate also
allows the SOC effects of the substrate to be transferred to the magnetic 3d
transition metal dimer. 
We have confirmed this by further calculations in which the SOC of the
substrate and the dimer atoms were manipulated separately. Enhancing the
SOC for a 
Co dimer leads primarily to an increase of the on-site
anisotropy $K_{2,2}$. However, enhancing the SOC for the Pt substrate
leads to a 
strong increase in the anisotropy $K_{2,1}$ as well as to a larger
difference in the individual torques on the two 
Co atoms, reflecting an increase of the DM interaction
(Fig.~\ref{fig:Co_MCA}). This behaviour is in line with Levi's model
of the indirect DM interaction between two spin moments \cite{Levi69},
which is mediated by nearby atoms. As a consequence, the magnitude of
the DM interaction is essentially determined by the SOC strength of the
neighbouring atoms.

\begin{figure}
\includegraphics[width=60.mm,angle=0,clip]{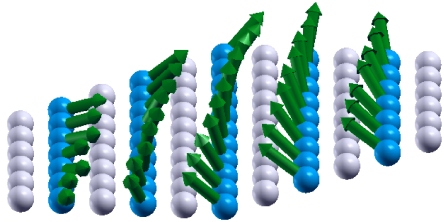}
\caption{\label{fig:FePt_2D} Magnetic structure of a 91-atom FePt ($2
  \times 1$) 
  2D alloy cluster deposited on the Pt(111) surface determined via MC
  simulation for $T = 0$K. The
  arrows denote the orientation of the magnetic moments associated with
  the iron atoms. }  
\end{figure}

The effect of anisotropic exchange is even more spectacular in magnetic
alloy nanoclusters, where magnetic atoms are separated by 
non-magnetic atoms with large SOC. This is demonstrated by our
calculations for a FePt ($2 \times 1$) 2D alloy cluster deposited on the
Pt(111) surface pertinent to an experimental investigation by Honolka et
al.~\cite{HLK+08}.
Ab-initio torque calculations show the clear trend that the magnetic 
structure of this cluster is non-collinear. 
Here we present the results of Monte Carlo simulation 
based on an effective Heisenberg model with the isotropic and DM
exchange coupling parameters  calculated in 
accordance to Ref. \cite{USPW03}. 
Fig.~\ref{fig:FePt_2D} shows the non-collinear magnetic structure
obtained for $T = 0$K. 
The non-collinearity between the Fe chains is essentially caused by the
nearest neighbours Fe-Fe interchain DM interaction ($|\vec{D}| = 4.6$
meV) being of similar magnitude when compared to the isotropic exchange
interaction ($J = 8.8$ meV). 
Within a Fe chain, however, the DM interaction is more than one
order of magnitude smaller when compared to the isotropic exchange
leading only to a slight screwing of the Fe magnetic moments along the chain.
Thus, we see that the FePt cluster's non-collinear structure is created 
by an  enhanced  DM coupling between Fe moments mediated by Pt atoms having
large SOC, on the one hand side, while the isotropic exchange between these
atoms separated by non-magnetic Pt is small enough to make both these
couplings comparable.

In summary our investigation of deposited transition metal dimers  as
simple but realistic model systems has
demonstrated that ab-initio magnetic torque calculations 
enable the
impact of SOC on the magnetic interactions within nanostructures to be
monitored in a very detailed way revealing subtle anisotropic
effects. 
The analysis of directly calculated electronic structure
quantities within a framework based on a Heisenberg model Hamiltonian
gives further insight and  identifies the role of the various
contributions and also the limitations of such models.
For the 
Fe and Co dimer systems studied here the DM interaction was found to be
pronounced 
owing primarily to the SOC of the substrate. Moreover it leads to
non-collinear magnetic configurations of the dimers in spite 
of the pronounced ferromagnetic coupling and out-of-plane
anisotropy. These magnetically anisotropic interactions also have a
profound effect in larger clusters and we have demonstrated this
explicitly with a study of a deposited FePt cluster. In particular we
infer that the magnetic structure around the edges of   
magnetic nanoparticles is likely to be significantly affected.

\subsection*{ Acknowledgement}

Financial support by the {\em Deutsche Forshungsgemeinschaft} within the
framework of the priority program {\em (DFG-Schwerpunktprogramm 1153)
  Clusters at Surfaces: Electron Structure and Magnetism} is gratefully
acknowledged.


\end{document}